%% file: ewsn-full.tex
\documentclass[10pt,emptycopyrightspace]{ewsn-proc}

\usepackage{graphicx}
\usepackage{balance}
\usepackage{comment}

\usepackage{orcidlink}

\usepackage{textcomp} 
\usepackage[nospace,noadjust]{cite} 
\usepackage{amsmath,amssymb,amsfonts} 
\usepackage{mathrsfs} 
\usepackage[mathscr]{euscript} 
\usepackage{epstopdf} 
\usepackage{xcolor}
\usepackage{pgfplots}
\pgfplotsset{compat=newest} 
\usepackage{tikz}
\usetikzlibrary{plotmarks}
\usepackage[normalem]{ulem} 
\usepackage{gensymb} 
\usetikzlibrary{intersections,backgrounds, patterns}
\usepackage{array}
\usepackage[colorinlistoftodos]{todonotes} 
\usepackage[nolist]{acronym} 
\usepgflibrary{arrows}
\usepackage{multirow} 

\usepackage{soul}

\usepackage[cmintegrals]{newtxmath} 
\usepackage[font=footnotesize]{subcaption}
\usepackage[font=footnotesize]{caption}

\numberofauthors{1}
\author{
\alignauthor Luisa Schuhmacher \orcidlink{0000-0003-3293-7933}, Sofie Pollin \orcidlink{0000-0002-1470-2076}, and Hazem Sallouha \orcidlink{0000-0002-1288-1023} \\
    \affaddr{Department of Electrical Engineering (ESAT) - WaveCoRE, KU Leuven}\\
    \email{\normalsize	{\{luisa.schuhmacher, sofie.pollin, hazem.sallouha\}@esat.kuleuven.be}}
}

\title{ecoBLE: A Low-Computation Energy Consumption Prediction Framework for Bluetooth Low Energy
}

\begin{document}

\maketitle
\input{Acro}

\begin{abstract}
    \input{abstract}
\end{abstract}

\section{Introduction}
    \input{introduction}

\section{Related Work}
    \label{sec:related_work}
    \input{relatedwork}

\section{Proposed Framework and Training Procedure}
    \label{sec:model_training}
    \input{model_training}
    \subsection{Proposed Framework}
        \label{sec:model}
        \input{model}

    \subsection{Training Procedure}
        \label{sec:training}
        \input{training}

\section{Experimental Setup and Dataset Collection}
    \label{sec:dataset}
    This section presents the experimental setup used for the dataset collection as well as the processing of the dataset.
    \input{dataset}

\section{Performance Evaluation}
    \label{sec:performance}
    \input{performance}

    \subsection{Performance Metrics}
        \label{sec:metric}
        \input{metric}

    \subsection{Results}
        \label{sec:results}
        \input{results}

\section{Conclusion}
    \label{sec:conclusion}
    \input{conclusions}

\section{Acknowledgment}
The present work has received funding from the European Union's Horizon 2020 Marie Skłodowska Curie Innovative Training Network Greenedge (GA. No. 953775). The work of Hazem Sallouha was funded by the Research Foundation – Flanders (FWO), Postdoctoral Fellowship No. 12ZE222N.

\balance
\bibliographystyle{abbrv}
\bibliography{references}
\end{document}

%% file: Acro.tex
\begin{acronym}

\acro{BLE}{Bluetooth Low Energy}
\acro{IoT}{Internet of Things}
\acro{ANNs}{Artificial Neural Networks}
\acro{LSTM}{Long Short-Term Memory}
\acro{LSTMP}{Long Short-Term Memory Projection}
\acro{MAPE}{Mean Absolute Percentage Error}
\acro{RNNs}{Recurrent Neural Networks}
\acro{GRU}{Gated Recurrent Unit}
\acro{SVR}{Support Vector Regression}

\end{acronym}

%% file: abstract.tex
Bluetooth Low Energy (BLE) is a de-facto technology for Internet of Things (IoT) applications, promising very low energy consumption. 
However, this low energy consumption accounts only for the radio part, and it overlooks the energy consumption of other hardware and software components. Monitoring and predicting the energy consumption of IoT nodes after deployment can substantially aid in ensuring low energy consumption, calculating the remaining battery lifetime, predicting needed energy for energy-harvesting nodes, and detecting anomalies. In this paper, we introduce a Long Short-Term Memory Projection (LSTMP)-based BLE energy consumption prediction framework together with a dataset for a healthcare application scenario where BLE is widely adopted. Unlike radio-focused theoretical energy models, our framework provides a comprehensive energy consumption prediction, considering all components of the IoT node, including the radio, sensor as well as microcontroller unit (MCU). Our measurement-based results show that the proposed framework predicts the energy consumption of different BLE nodes with a Mean Absolute Percentage Error (MAPE) of up to 12\%, giving comparable accuracy to state-of-the-art energy consumption prediction with a five times smaller prediction model size.

%% file: introduction.tex
The increasing trend in the number of wirelessly connected devices triggers considerable environmental and economic concerns, citing energy efficiency \cite{sallouha2017ulora} and sustainability of wireless networks \cite{wang2021evolution, ICTenergy}. In fact, the ICT sector, which is considerably driven by mobile networks and \emph{\ac{IoT}}, contributes to around 10\% of the total energy consumption worldwide, translating into 3\% of worldwide carbon footprint \cite{ICTenergy}. With the rapid increase of IoT networks and the belief that digitalization and wireless connectivity will play a key role in decreasing carbon emissions, and therefore tackling climate change \cite{ye2021using}, it is necessary to ensure low energy consumption of the IoT network itself \cite{sallouha2017ulora}.

\emph{\ac{BLE}} is emerging as one of the main drivers of \ac{IoT} technologies, promising low-energy and low-cost solutions for a great variety of short-range applications, including wearable devices, healthcare, and home automation to name a few \cite{raza2015bluetooth, chen2020wearable}. 
However, despite the fact that the \ac{BLE} protocol is inherently designed to be energy efficient, this efficiency is mostly focused on the radio frequency (RF) part of the IoT node and is not necessarily guaranteed on the network level as it highly depends on the scale of network deployment and traffic-demand, as well as the type of hardware used \cite{shan_optimal_2016, siekkinen_how_2012, kindt_energy_2020}. 

Energy consumption monitoring and prediction are part of the main research areas aiming at reducing wireless networks' carbon footprint, which also include areas focusing on energy-harvesting and battery-life monitoring. 
Energy consumption prediction models can be broadly categorized into models based on statistical approaches and models based on \emph{\ac{ANNs}} \cite{farouk_survey_2018, shao_prediction_2020, kong_short-term_2019}. Recent reports showed that approaches based on \ac{ANNs} outperform conventional statistical models in time-series problems \cite{savi_short-term_2021, sallouha2021aerial}. In particular, the \emph{\ac{LSTM}} \cite{hochreiter_long_1997} architecture is leading in terms of accuracy when dealing with prediction tasks on time-series datasets \cite{kong_short-term_2019}. Nonetheless, a critical shortcoming of the \ac{LSTM} architecture is its complexity as it incorporates multiple matrix operations, resulting in often large sequential model size. This makes it unsuitable for the current trend in \ac{IoT}, where machine learning models are implemented on computationally-limited edge devices \cite{hu_dynamic_2019}. To address the complexity of the conventional \ac{LSTM}-based models, a modification to the \ac{LSTM} architecture called \emph{\ac{LSTMP}} was introduced in \cite{sak_long_2014}. The authors showed that the \ac{LSTMP} architecture could reduce the overall size of the conventional \ac{LSTM} without suffering from degraded accuracy.

In this paper, we aim to leverage the computationally- and memory-efficient \ac{LSTMP} architecture, advocating for Green AI option \cite{schwartz_green_2019} by considering the model size in performance evaluation. In particular, the contribution of this paper is twofold. First, we conducted extensive measurements to collect an energy consumption dataset from \ac{BLE} nodes in a scenario resembling three healthcare \ac{BLE} sensors and a central \ac{BLE} device acting as a gateway. 
The dataset is made publicly available\footnote{https://doi.org/10.48804/XVOUAH}. To the best of our knowledge, such a dataset does not yet exist publicly. Secondly, we introduce a novel, low-computation \ac{LSTMP}-based energy consumption prediction framework trained on the said dataset. The proposed framework is five times smaller than its state-of-the-art counterpart \cite{alonso_single_2020} while staying on par in terms of prediction accuracy, making it more suitable for energy-, memory- and computation-restrained \ac{IoT} edge devices. 
Our proposed framework is made open-source\footnote{https://gitlab.kuleuven.be/networked-systems/public/energy-consumption-prediction-for-ble}, allowing it to act as a baseline for future research.

The rest of this paper is organized as follows: Section \ref{sec:related_work} reviews related work in the field of energy consumption prediction and energy consumption models for \ac{BLE} and beyond. We introduce our own energy consumption prediction framework and our training procedure in Section \ref{sec:model_training}. In Section \ref{sec:dataset}, we state our experimental setup for collecting the dataset used for training. Section \ref{sec:performance} illustrates the results we got from our framework. We conclude our work in Section \ref{sec:conclusion}.

%% file: relatedwork.tex
Several works in the literature have addressed energy prediction using machine-learning-based models; however, most targeted households \cite{farouk_survey_2018}. A survey on general energy consumption models has been given in \cite{farouk_survey_2018}, where several machine learning algorithms and their respective applications are summarized. In methods based on \ac{ANNs}, \ac{LSTM} models often take the lead for energy consumption prediction. Alonso \textit{et al.} \cite{alonso_single_2020} proposed an \ac{LSTM}-based framework and tested it on residential households’ energy consumption data, outperforming state-of-the-art models for this task. Their model is comparably complex \cite{alonso_single_2020} but showed its capability to generalize to data from unseen households while testing.

Regarding IoT networks, a few works have studied the energy prediction problem. Jeddou \textit{et al.} \cite{jeddou_power_1} studied the energy consumption prediction for the MQTT and COAP protocols. They use a simple linear regression model and achieve sufficient results. However, these results are only valid for data that follow a linear trend. Theoretical energy consumption models like the concise energy model for the \ac{BLE} protocol developed by \cite{kindt_energy_2020} can be used for energy consumption prediction. However, to use this model, several fine-grained measurements of the current drained as well as the duration of multiple parts of a \ac{BLE} application needs to be taken, which is specific to the used hardware and the software used to program the node, making it therefore tedious to adopt. Furthermore, it only comprises the radio, leaving software- and other hardware-components out.

In summary, while several works in the literature have focused on the field of energy consumption forecasting, the majority of them targeted households \cite{farouk_survey_2018}, and only a few research reports have focused on energy consumption prediction for \ac{IoT}. To the best of our knowledge, particularly, the comprehensive energy consumption prediction for \ac{BLE} is still an open research question.

%% file: model_training.tex
In this section, we introduce our \ac{LSTMP}-based prediction framework and detail the training procedure we followed.

%% file: model.tex
\begin{figure}
    \centering
    \includegraphics[width=0.45\textwidth]{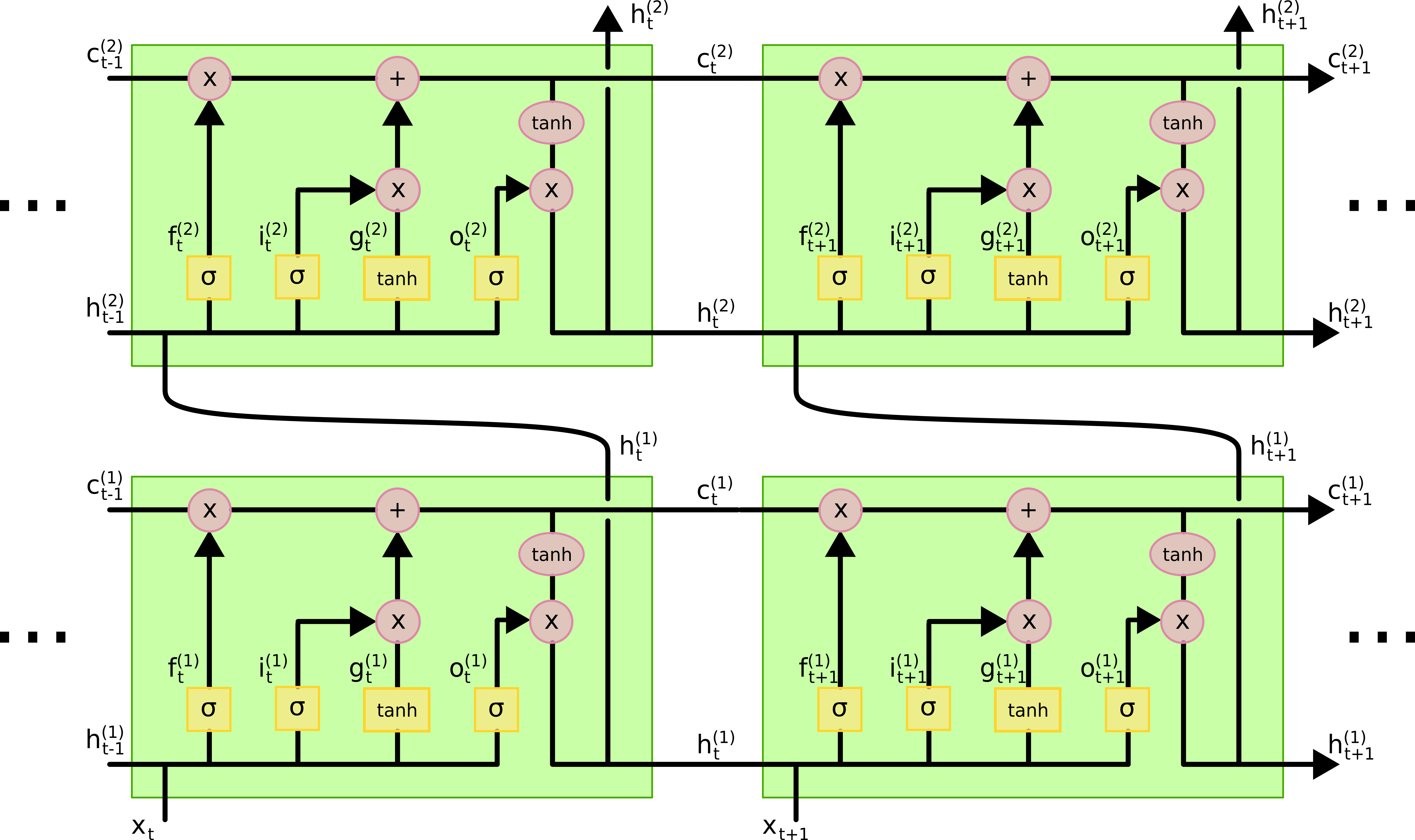}
    \caption{Visualization of our two-layer \ac{LSTMP} architecture. Subscript $t$ denotes the time-step, while superscript $(l)$ indicates the layer. $h^{(l)}_t/h^{(l)}_{t+1}$ are depicted twice to indicate that they are fed as input into the \ac{LSTMP} framework at the next time-step and simultaneously act as the final output of the current time-step. Note that the learnable parameters are different for the two layers (vertical direction) but the same for each time-step (horizontal direction).}
    \label{fig:lstm}
    \vspace{-0.5cm}
\end{figure}

Our framework consists of an ANN with two layers of \ac{LSTMP}, aiming to predict the power consumption of \ac{BLE} nodes. The architecture of our \ac{LSTMP}-based ANN is visualized in Figure \ref{fig:lstm} for input from time-steps $t$ and $t+1$. 

The only difference between a conventional \ac{LSTM} layer and an \ac{LSTMP} layer is a projection matrix $W_{hr}$ which reduces the size of the hidden state vector, $h_t^l$ in Figure \ref{fig:lstm}, from the hidden size to a predefined projection size. We set the projection size to be the prediction length, directly decoding the features gathered by the gates to the predicted energy consumption. 
Note that like this, we erase the necessity for an additional linear layer after the \ac{LSTMP} to obtain the desired output. At the same time, the sizes of all the matrices used to compute the hidden state 
become less compared to the conventional \ac{LSTM} architecture, leading to a decreased overall architecture size.

In the next subsection, we will elaborate on the training procedure. We use the symbols as described in Table \ref{tab:symbols}.
\begin{table}[]
    \centering
    \small
    \caption{Symbols used throughout this paper.}
    \begin{tabular}{|l|l|l|}
        \hline
        Symbol & Meaning in general & Meaning in Fig. \ref{fig:lstm} \\
        \hline
        \hline
        $T$ & Length of sequence & $t\in\{0,...T-1\}$ \\
        & input vectors given to & \\
        & the framework & \\
        \hline
        $H_{in}$ & Size of one input vector & Size of $x_t$ \\
        & given to the framework & \\
        \hline
        $H_{cell}$ & Hidden size for cell & Size of $c^{(l)}_t,f^{(l)}_t,$ \\
        & state and hidden state, & $i^{(l)}_t,g^{(l)}_t,o^{(l)}_t$ \\
        & output size of cell state & \\
        \hline
        $H_{out}$ & Output size of hidden & Size of $h^{(l)}_t$ \\
        & state and actual output & \\
        \hline
    \end{tabular}
    \label{tab:symbols}
\end{table}
\renewcommand{\arraystretch}{1}

%% file: training.tex
In order to select the best performing \ac{LSTMP} architecture, we trained multiple two-layer \ac{LSTMP} networks with different hidden sizes $H_{cell}$ and different output sizes $H_{out}$. Our objective is to realize a model that can be implemented directly on a sensor with limited computational power instead of the cloud. An example case is predicting the energy an energy-harvesting node will need in a given amount of time-steps in the future. As a result, we aim at minimizing the size and the computation requirements of the proposed framework.
Therefore, we search through several values of $H_{cell}$ to determine a small but accurate framework. 

We further vary $H_{out}$ to let the framework predict the energy consumption for different amounts of future time-steps. We want to investigate how far we can let the framework predict without substantially degrading the target accuracy. Table \ref{tab:sizes} shows the combinations of values for $H_{cell}$ and $H_{out}$ used during training.
\begin{table}[]
    \centering
    \small
    \caption{Hidden sizes $H_{cell}$ and output sizes $H_{out}$ used in training.}
    \begin{tabular}{|l|l|}
        \hline
        $H_{cell}$ & $H_{out}$ \\
        \hline
        \hline
        2 & 1 \\
        \hline
        4 & 1 \\
        \hline
        8 & 1 \\
        \hline
        16 & 1 \\
        \hline
        32 & 1, 2, 3, 10, 15, 50 \\
        \hline
        64 & 1, 2, 3, 10, 15, 50 \\
        \hline
        650 & 1 \\
        \hline
    \end{tabular}
    \label{tab:sizes}
\end{table}
In order to evaluate how many time-steps of 1 ms in the future our framework can predict, we started with the simplest task of predicting one step in the future and evaluated the performance for the different values of $H_{cell}$. 
Subsequently, we fixed $H_{cell}\in\{32,64\}$, as they yielded the best prediction accuracy (cf. Section \ref{sec:performance}), without further improvement for bigger $H_{cell}$, and subsequently, trained the models with different $H_{out}$.

%% file: dataset.tex
\subsection{Raw Dataset}
To train and test our framework, we set up an experiment to build a dataset. The experiment was designed to imitate a healthcare application where a patient is equipped with three sensors, each of which has a \ac{BLE} radio node attached to provide wireless connectivity, sending information via \ac{BLE} to a central device. The first sensor, denoted by HR, reads the pulse of the patient at 1000 samples per second (sps), and the attached \ac{BLE} node then calculates the heart rate and sends it to a central node every five seconds. The second sensor, denoted by BT, reads the patient's body temperature and sends it via \ac{BLE} to the central node every five minutes. The last sensor, denoted by OS, reads the oxygen saturation in the patient's blood, and the corresponding \ac{BLE} node sends it to the central node every second. Since the actual data read by the sensor has no influence on the proposed framework, we used dummy data generated by the boards' MCU for HR and OS, while air temperature sensor readings were used for BT in our dataset collection. 

\begin{figure}
    \centering
    \includegraphics[width=0.35\textwidth]{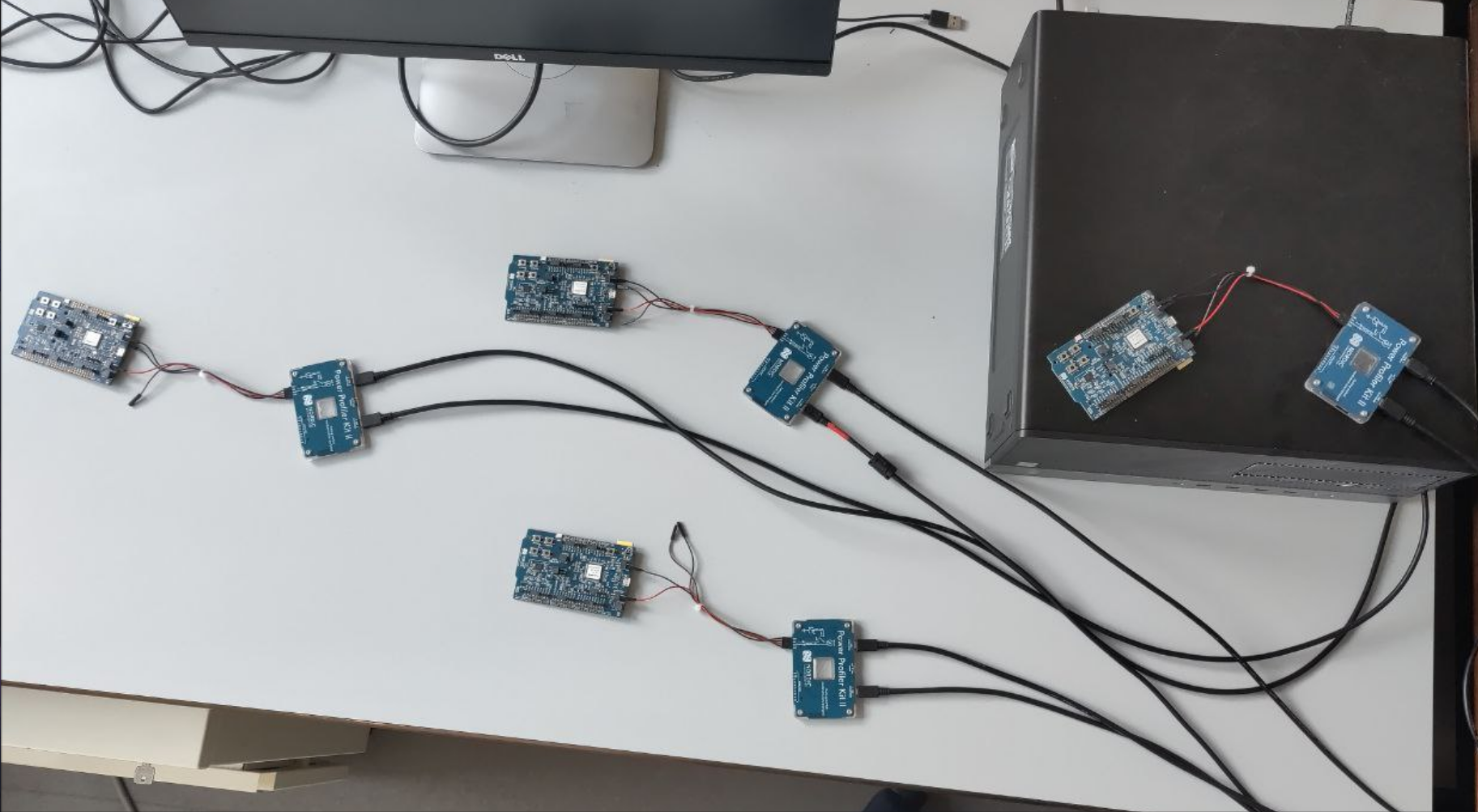}
    \caption{Experimental setup. Each \ac{BLE} board is connected to and powered by a Power Profiler, which sends the current measurements to a computer via cable.}
    \label{fig:setup}
\end{figure}

In this experimental setup, we used Nordic's nRF52 Development Kit \cite{nrf52dk} with their nRF52832 System-on-Chip \cite{nrf52832soc}, running Nordic's nRF Connect Software Development Kit \cite{nrfconnectsdk}, which includes Zephyr RTOS \cite{zephyrproject}. The code we flashed the nodes with is based on Zephyr's code samples \cite{bluetooth:samples}. Each \ac{BLE} node was also connected to Nordic's Power Profiler Kit II \cite{powerprofilerkit2} to provide accurate current measurements. We adapted an unofficial python API for the power profiler \cite{ppk2pythonapi} to automate the measurements of our experiment and collected current measurements for three hours with a 100 kHz sampling rate for all four nodes. Figure \ref{fig:setup} shows our experimental setup. We let the experiments run in an office with people entering and leaving and phones connected to WiFi to replicate, to some extent, a hospital room environment.
Besides timestamp and current, four time-independent features are included, which are summarized in Table \ref{tab:features}.
\begin{table}[]
    \centering
    \small
    \caption{Features aside from the current drained in our dataset. "hr" denotes the heart rate sensor, "bt" denotes the body temperature sensor, and "os" denotes the oxygen saturation sensor.}
    \begin{tabular}{|l|l|}
        \hline
        Name & Explanation/possible values \\
        \hline
        \hline
        Role & "master" or "slave" \\
        \hline
        Nr. connections & \# nodes this node is connected to \\
        \hline
        Transmission rate (pps) & \# packets sent per second \\
        \hline
        Packet size (B) & \# bytes sent in a transmission \\
        \hline
        Node & "central", "hr", "bt" or "os" \\
        \hline
    \end{tabular}
    \label{tab:features}
\end{table}

\subsection{Dataset Preprocessing}
To create the dataset with which we trained our framework, we down-sampled the measurements by a hundredfold to 1 kHz by using the sums of the measurements as well as the maximums. The motivation for down-sampling is, on the one hand, to reduce the size of the dataset and, on the other hand, to decrease the number of time-steps the \ac{LSTMP} framework needs to process to gather meaningful information about past connections of the \ac{BLE} node. 
We selected the down-sampling by a hundredfold and made sure to keep
connection events visible in the dataset without mixing with the sleeping events. By taking the sum as well as the maximum values, we capture both long events and short peaks. 

Figure \ref{fig:pred_ex} shows sample traces for each node, where they are compared to one of our models' predictions. The other features are time-invariant and do not change for a given node, so down-sampling them is straightforward. The structure of this down-sampled dataset 
can be seen in Table \ref{tab:dataset_downsampled}.
\begin{table*}[]
    \centering
    \small
    \caption{Down-sampled dataset with timestamps, the sum of currents drained, maximum current drained, and time-independent features.}
    \begin{tabular}{|c|c|c|c|c|c|c|c|}
        \hline
        Timestamp (\micro s) & Sum of currents (\micro A) & Max. of currents (\micro A) & Role & Nr. & Transm. & Packet & Node \\
        &&&& conn. & rate (pps) & size (B) & \\
        \hline
        \hline
        1725600000 & 105453.918095073 & 4038.58944143533 & master & 3 & 0 & 0 & central  \\
        1725601000 & 91505.1027793711 & 2783.95385863702 & master & 3 & 0 & 0 & central  \\
        1725602000 & 264778.571039428 & 11329.6786862516 & master & 3 & 0 & 0 & central  
    \end{tabular}
    \label{tab:dataset_downsampled}
\end{table*}

%% file: performance.tex
In this section, we introduce the evaluation metrics used in this work and assess the performance of our framework on the preprocessed dataset. We use as a baseline the state-of-the-art \ac{LSTM} framework introduced in \cite{alonso_single_2020}. Both our framework and the baseline framework are capable of predicting the energy consumption for any of the \ac{BLE} nodes considered in this work. As input, they receive samples from our down-sampled dataset (Table \ref{tab:dataset_downsampled}). We use Pytorch Lightning \cite{Falcon_PyTorch_Lightning_2019} for implementation and training. Moreover, we leverage Pytorch Forecasting \cite{Beitner_PyTorch_Forecasting_2020} for encoding the features in our dataset and creating a Pytorch dataset for training. We use a 70-15-15 split of the dataset for training, validating, and testing the frameworks, respectively. In particular, we take the first 70\% of the measurements of each node as a training set, the next 15\% of each node as a validation set, and the last 15\% of each node as a test set, ensuring that each set has the same amount of samples for each node.

%% file: metric.tex
To study the performance of our proposed framework, we investigate the model accuracy and motivated by the spirit of Green AI, also the model size. To evaluate the accuracy, we adopt the \ac{MAPE} between the prediction $\hat{y}$ and the actual consumption $y$ on the test set, which can be written as
$$\text{MAPE}(y, \hat{y})=\frac{1}{n}\sum_{i=0}^n\frac{|y_i-\hat{y}_i|}{\max(\epsilon, |y_i|)},$$
where $n$ is the number of samples, and $\epsilon$ is a very small, strictly positive number (to prevent division by zero). \ac{MAPE} is a well-established metric in the time series prediction problems and can be used to compare performances independent of the scaling or the dataset trained on. 
To assess the model size, we report the number of parameters for our framework as well as the baseline framework.

%% file: results.tex
This subsection presents the results of our framework training.
\subsubsection{Hidden Size Search}
\begin{figure}[]
    \begin{minipage}{0.49\linewidth}
        \centerline{\includegraphics[width=\linewidth]{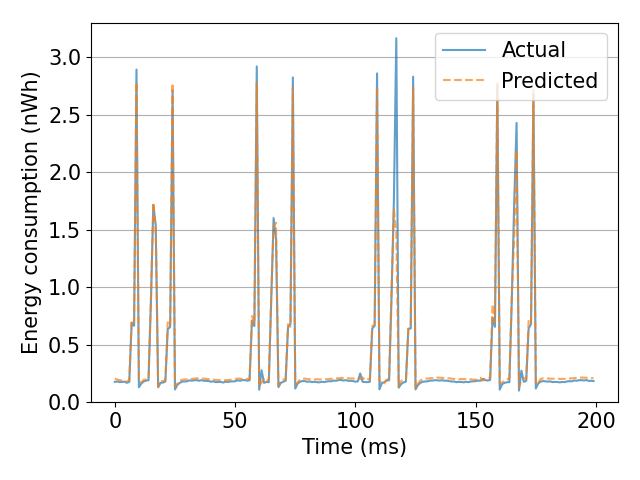}}
        \caption*{a) Central}
    \end{minipage}
    \hfill
    \begin{minipage}{0.49\linewidth}
        \centerline{\includegraphics[width=\linewidth]{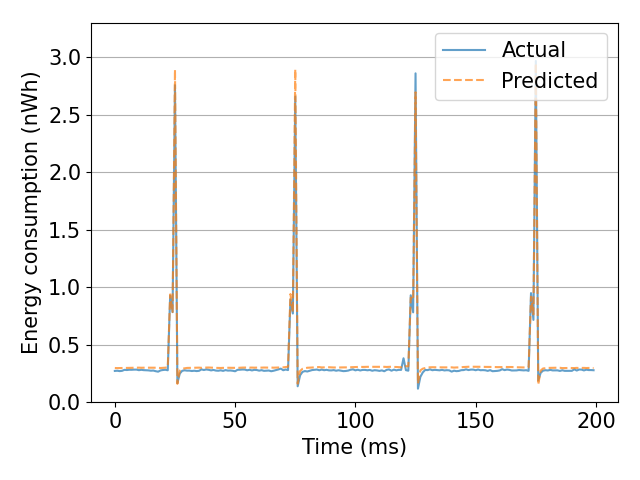}}
        \caption*{b) HR}
    \end{minipage}
    \begin{minipage}{0.49\linewidth}
        \centerline{\includegraphics[width=\linewidth]{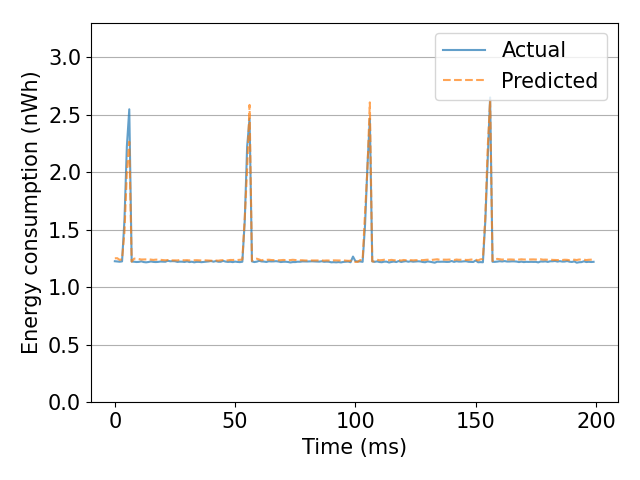}}
        \caption*{c) BT}
    \end{minipage}
    \hfill
    \begin{minipage}{0.49\linewidth}
        \centerline{\includegraphics[width=\linewidth]{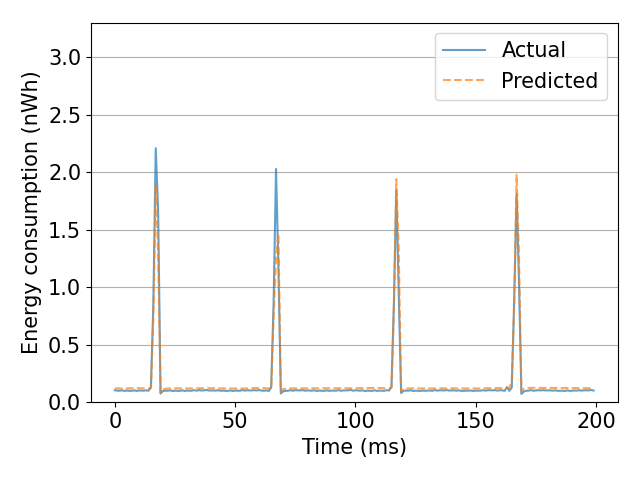}}
        \caption*{d) OS}
    \end{minipage}
    \caption{Traces of energy consumption from the test set (blue) for each node and the prediction made by the same model, which takes 50 time-steps as input sample and predicts one step in the future (orange). The time period we use to calculate the energy consumption is 1 ms. Note that BT is the only node that uses actual sensor readings and, therefore, also powers the sensor.}
    \label{fig:pred_ex}
\end{figure}

First, we trained our proposed framework with different values of $H_{cell}$ to predict one time-step into the future. A sample trace from the dataset with the prediction of a model trained from our framework can be seen in Figure \ref{fig:pred_ex}. It can be seen that the predictions are accurate for all four nodes considered in this work, regardless of their type or role.

\begin{figure}[]
    \centerline{\includegraphics[width=0.4\textwidth]{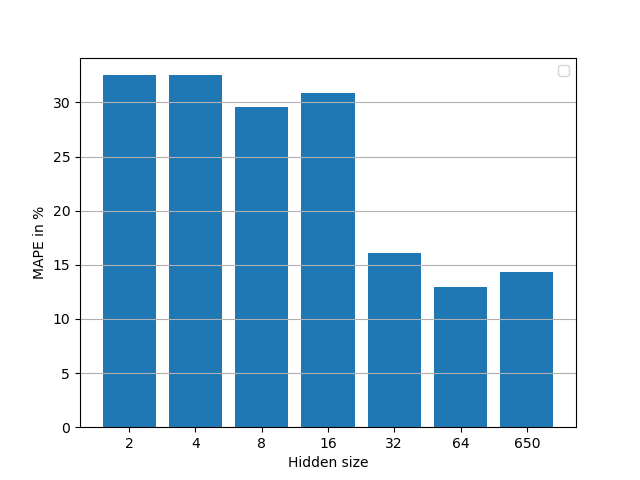}}
    \caption{\ac{MAPE} in \% for our framework with various hidden sizes $H_{cell}$. \linebreak A lower \ac{MAPE} indicates better accuracy.}
    \label{fig:mape_hs}
\end{figure}

Figure \ref{fig:mape_hs} shows the \ac{MAPE}s of the proposed LSTMP framework achieved on the test set with different values $H_{cell}$. The figure illustrates that when $H_{cell}\leq16$, the LSTMP-based ANN shows poor performance, achieving only a \ac{MAPE} of around 30\%. For the case with $H_{cell}=32$ and $H_{cell}=64$, the \ac{MAPE} falls under the 20\% mark, to 16.13\% and 12.96\%, respectively, indicating high prediction accuracy. Increasing $H_{cell}$ further does not yield any improvement. Figure \ref{fig:mape_hs} also shows that a huge model with $H_{cell}=650$ achieved a \ac{MAPE} of 14.38\% as well, providing comparable performance to the LSTMP-based models with $H_{cell}=32$ and $H_{cell}=64$. 

\subsubsection{Prediction Length}
Following the results from Figure \ref{fig:mape_hs}, we subsequently explore how many steps in the future our \ac{LSTMP}-based models with $H_{cell}=32$ and $H_{cell}=64$ can predict and compare it to the baseline \ac{LSTM}. The different $H_{out}$'s used and the meaning behind them in the context of our \ac{BLE} application are stated in Table \ref{tab:proj_meanings}. 
\begin{table}[]
    \centering
    \small
    \caption{Output sizes $H_{out}$ and their meaning for the dataset.}
    \begin{tabular}{|l|l|}
        \hline
        Output size & Meaning \\
        \hline
        \hline
        1, 2 & Next step(s) of the connection / sleep \\
        \hline
        3 & Whole connection event \\
        \hline
        10 & Two connection events (central only) \\
        \hline
        15 & Three connection events (central only) \\
        \hline
        50 & Connection event and sleep \\
        \hline
    \end{tabular}
    \label{tab:proj_meanings}
\end{table}
Note that only the central node has three connections. 
The sequence length $T$, which the \ac{LSTMP} models see in the past, is fixed at 50 time-steps, translating into 50 ms. We chose $T = 50$ empirically because 50 ms of former energy consumption are needed to see the preceding connection with the parameters we have set in our \ac{BLE} application. Only with this information can a prediction be made about when the next connection will occur.

\begin{figure}[]
    \centerline{\includegraphics[width=0.4\textwidth]{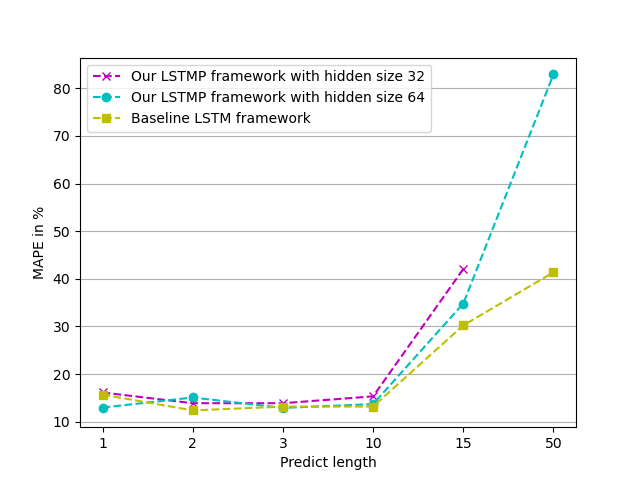}}
    \caption{\ac{MAPE} in \% for our framework with hidden size $H_{cell}=32$ and $H_{cell}=64$ predicting various amounts of steps in the future in comparison with the baseline \ac{LSTM} model from \cite{alonso_single_2020}. A lower \ac{MAPE} indicates better accuracy.}
    \label{fig:mape_pl}
\end{figure}

\begin{figure}[]
    \centerline{\includegraphics[width=0.4\textwidth]{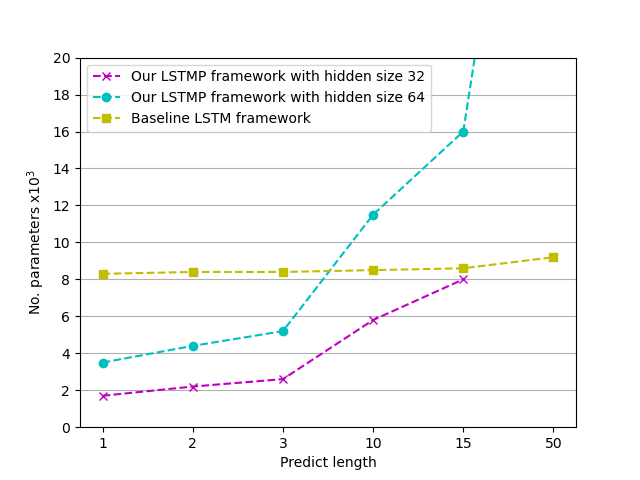}}
    \caption{Number of model parameters for our framework with hidden size $H_{cell}=32$ and $H_{cell}=64$ predicting various amounts of steps in the future in comparison with the baseline \ac{LSTM} model from \cite{alonso_single_2020}. For the sake of visualization, the number of parameters for our framework with $H_{cell}=64$ predicting 50 steps in the future cannot be seen.}
    \label{fig:size_pl}
\end{figure}

The results from our performance evaluation are shown in Figures \ref{fig:mape_pl} and \ref{fig:size_pl}. Figure \ref{fig:mape_pl} shows the performance in terms of the \ac{MAPE} score. It can be seen that our framework can predict up to 10 steps in the future without substantially losing accuracy for both $H_{cell}$, exhibiting similar performance to the baseline. When predicting 15 or 50 steps in the future, the \ac{MAPE} for both our framework and the baseline exhibits a noticeable jump to above 30\%, indicating that they do not predict accurately anymore.

Figure \ref{fig:size_pl}, on the other hand, shows the number of parameters of the baseline framework and our framework for different prediction lengths. It can be seen that for the number of prediction steps with an acceptable \ac{MAPE}, i.e., for a prediction length $H_{out}\leq10$, our framework with $H_{cell}=32$ is substantially smaller than the baseline. For a one-step prediction, it has about five times fewer parameters. It can also be seen, however, that the model size increases quicker for longer prediction lengths than the baseline, indicating that the baseline model is the better choice for prediction tasks where a substantial amount of future steps need to be predicted at once.

Notice that we did not train our framework with $H_{cell}=32$ and $H_{out}=50$ because for an \ac{LSTMP}, the constraint $H_{out}<H_{cell}$ must be satisfied.

\subsubsection{Performance for each Node}
Until now, we have only looked into our framework's performance for all four aggregated nodes with different roles. In the following, we investigate the performance of a model from our framework with $H_{cell}=64$ and $H_{out}=1$. Figure \ref{fig:pred_ex} shows sample traces with predictions from our model for all four nodes. It can be seen in the figure that the model accurately adapts to the different energy consumption values in the sleep phase and that it is also able to predict the three connection events in the case of the central node, while it only predicts one connection event for each of the other three nodes. Moreover, we calculated the \ac{MAPE}s for each node separately, as shown in Table \ref{tab:mapes_separately}.
\begin{table}[]
    \centering
    \small
    \caption{\ac{MAPE}s achieved by a model on the current consumption data of each node separately. HR denotes the heart rate sensor, BT denotes the body temperature sensor, and OS denotes the oxygen saturation sensor.}
    \begin{tabular}{|l|c|c|c|c|}
        \hline
        \textbf{Node} & Central & HR & BT & OS \\
        \hline
        \textbf{\ac{MAPE}} & 22.66\% & 12.36\% & 3.08\% & 13.72\% \\
        \hline
    \end{tabular}
    \label{tab:mapes_separately}
\end{table}
It can be seen that the performance for HR and OS with a \ac{MAPE} of 12.36\% and 13.72\%, respectively, are about the same as the \ac{MAPE} for the whole dataset, which is 12.96\%. For the central node, the performance drops to 22.66\%, which can be explained by the fact that thrice as many connection events happen as for the other nodes. During the connection events, unpredictable peaks in the current consumption occur, negatively affecting our model's accuracy. An example of this can be seen in Figure \ref{fig:pred_ex} a). On the BT data, however, the performance increases to a \ac{MAPE} of only 3.08\%, indicating highly accurate predictions for this specific node, which can be attributed to this node's relatively low transmission rate.

\subsubsection{Positioning our Model with Respect to Simple Store and Copy}
Since the dataset used in this paper is repetitive, we compared our model to a simple store and copy approach. We chose a copying window of 50 ms, representing the connection interval. Our experimental results showed that copying provides better performance for all nodes combined. However, this copying mechanism requires perfect knowledge of the exact value of the connection interval. In order to position our model, we trained it with different sequence lengths that are shorter and longer than the Connection Interval, representing a case without knowledge of the exact connection interval. Our experiments showed that using sequence lengths longer than the Connection Interval does not impact the accuracy of our model, and for the case of a shorter sequence length, the performance drops a few percentages to 22.08\%. Further investigation is required with different training sequences, which is left for future work.

%% file: conclusions.tex
We have introduced a fine-grained energy consumption dataset for a healthcare-resembling application with four \ac{BLE} nodes. The dataset entails multiple features of the nodes and fine-grained current measurements and is made publicly available. Moreover, an accurate energy consumption prediction framework has been proposed. The proposed framework maintains an acceptable prediction accuracy with \ac{MAPE}s below 25\%, comparable to a state-of-the-art prediction model while reducing the model complexity to a fifth. In addition, we showed that the proposed framework is able to predict 10 time-steps in the future with a \ac{MAPE} lower than 20\%. Furthermore, we demonstrated that the proposed framework keeps a \ac{MAPE} lower than 25\% when examined on \ac{BLE} nodes resembling a critical healthcare scenario separately, with different numbers of connections and transmission intervals.